# Extensile to contractile transition in active microtubule-actin composites generates layered asters with programmable lifetimes.

John Berezney[1], Bruce L. Goode[2], Seth Fraden[1], Zvonimir Dogic[3,4,1]

[1]Department of Physics, Brandeis University, Waltham MA 02454
[2]Department of Biology, Brandeis University, Waltham MA 02454
[3]Department of Physics, University of California at Santa Barbara, Santa Barbara, CA 93106
[3]Biomolecular Science and Engineering Program, University of California at Santa Barbara, Santa Barbara, CA 93106

**Abstract:** We study a reconstituted composite system consisting of an active microtubule network interdigitated with a passive network of entangled F-actin filaments. Increasing viscoelasticity of the F-actin network controls the emergent dynamics, inducing a transition from turbulent-like flows to bulk contractions. At intermediate F-actin concentrations, where the active stresses change their symmetry from anisotropic extensile to isotropic contracting, the composite separates into layered asters that coexist with the background turbulent fluid. Contracted onion-like asters have a radially extending microtubule-rich cortex that envelops alternating layers of microtubules and F-actin. The self-regulating layered organization survives aster merging events, which are reminiscent of droplet coalescence, and suggest the presence of effective surface tension. Finally, the layered asters are metastable structures. Their lifetime, which ranges from minutes to hours, is encoded in the material properties of the composite. Taken together, these results challenge the current models of active matter. They demonstrate that the self-organized dynamical states and patterns, which are evocative of those observed in the cytoskeleton, do not require precise biochemical regulation but can arise due to purely mechanical interactions of actively driven filamentous materials.

**Significance statement:** Active forces sculpt the forms of living things, generating robust, adaptable, and reconfigurable dynamical materials. Creating synthetic materials which exhibit comparable control over internally generated active forces remains a challenge. We demonstrate that active composite networks, collectively driven by the force-generating molecular motors, exhibit complex spatiotemporal patterns that are similar to those observed in cell biology. Amongst others, we describe how mechanical properties of the network enable spatial and temporal control of onion-like layered asters. Asters are reminiscent of liquid droplets, yet their self-regulating mechanism ensures their layered structure survives coalescence-like events. Our model system elucidates the essential role of passive elasticity in controlling the emergent non-equilibrium dynamics, while also establishing a robust experimental platform for engineering life-like materials.

**Introduction:** Cytoskeletal active stresses, which are collectively generated by thousands of microscopic molecular motors, empower living organisms with the capacity to reshape themselves and their environment. Precise control of these self-organized stresses in both space and time is essential for diverse biological processes including embryogenesis, intracellular transport, cell motility, and cytokinesis (1-3). Reconstituting such spatiotemporal dynamics in synthetic active materials remains a challenge, but one that is worth pursuing for several reasons. From a fundamental perspective, the development of experimental active matter systems provides an opportunity to test the emerging theories of internally driven non-equilibrium systems (4). In turn, such theoretical frameworks could enable a rigorous description of various dynamical architectures observed in biology (5, 6). From an applications perspective, active matter systems could provide an experimental platform for creating a new class of autonomous shape-changing materials that mimic sought-after properties of living organisms.

The above-described considerations inspired studies of structures and dynamics that emerge in collections of interacting motile agents, which are either chemically synthesized or biologically isolated. Several behaviors have been observed. Self-propelled Brownian particles which lack structural anisotropy, form dense condensates that coexist with a dilute gas and are reminiscent of equilibrium gas-liquid phase separation (7, 8). Motile objects with polar symmetry form dense flocks that exhibit directed sinuous motion (9). Anisotropic active agents, such as bacterial swimmers or molecular-motor powered extensile bundles, remain spatially homogeneous but generate turbulent-like dynamics and autonomous flows (10-12). Finally, cytoskeletal filaments and molecular motors also exhibit bulk contractions into dense solid-like materials (13-17). These advances provide a foundation for introducing biological functionality into materials science. However, several barriers remain. First, predicting the emergent self-organization and the bulk material properties from known microscopic dynamics remains a challenge (18, 19). For example, it is difficult to predict which combination of motor proteins and cytoskeletal filaments will exhibit extensile turbulent-like dynamics or contractile network collapse (20). Second, living organisms can rapidly assemble large-scale dynamical architectures, disassemble these same structures at a later predetermined time, and then reassemble the same constituents into other distinct structures. In contrast, engineering synthetic active materials with multiple dynamical states and controlling the switching between these states remains a challenge.

Motivated by these twin challenges, we developed a composite system of cytoskeletal active matter that merges an active microtubule (MT)-based extensile network powered by kinesin-1 motors and a passive network of entangled F-actin filaments. Such composites of active and passive filaments enable independent control of the magnitude of the active stresses, exclusively generated by the active MT component, and the passive viscoelasticity, dominated by the entangled network of F-actin filaments. This additional degree of freedom widens the accessible phase space, whose full exploration reveals diverse dynamical states and complex kinetic pathways. With increasing viscoelasticity, we observe a transition from extensile turbulent-like dynamics to bulk contraction. The transition regime is characterized by the coexistence of an extensile active fluid and locally contracted layered asters with liquid-like properties. However, unlike conventional droplets, the finite-lifetime structured asters exhibit a layered onion-like spatial structure that survives coalescence. Taking these behaviors together, the active MT and passive actin composites share certain features of seemingly distinct categories of active matter, including extensile and contractile networks, as well as active condensation.

**Composite MT-actin networks:** To prepare a composite network with tunable mechanical properties, we combined active MT bundles with a passive actin network whose viscoelasticity is well-controlled. G-actin polymerizes into microns-long semi-flexible filaments that, at high concentrations, form a viscoelastic entangled network with an elastic plateau at intermediate frequencies (**Fig. 1A**) (21, 22). With increasing F-actin concentration this elastic plateau ranges from 0.01 to 2 Pa, higher than the elasticity of the MT networks alone (23). Consequently, the composite's viscoelasticity is dominated by the F-actin component, effectively decoupling the network's passive mechanics from the active stresses that are only generated by the MT component. The previously used non-specific bundling agent, poly(ethylene glycol), induced coassembly of mixed MTs and F-actin bundles that lacked activity (24, 25). To produce a force-generating MT network we used protein regulator of cytokinesis (PRC1), a crosslinker that specifically bundles anti-parallel MTs, while still allowing for the kinesin-driven interfilament sliding (26, 27).

**Controlling initial conditions:** To ensure reproducible dynamics, we prepared samples with the same ATP-independent initial state. Both actin polymerization and reconfigurations of extensile MT networks are ATP-dependent processes, but they take place on different time scales. G-actin polymerizes over hours, while the active MT network rearranges on the time scale of seconds (23, 28). Upon simply mixing all the components, the MTs formed an extensile network within minutes, which thereafter generated substantial flows before G-actin could polymerize into F-actin. The ATP-powered MT-driven flow churned the polymerizing actin into various flow-dependent structures resulting in ATP-dependent initial conditions that violated our design constraints.

To overcome this challenge, we developed a two-step protocol. First, we prepared samples with the same initial configuration consisting of a fully polymerized F-actin network interdigitated with a passive MT network (**Fig. 1A, 1B**). Subsequently, we activated the MT network to initiate the non-equilibrium dynamics. To accomplish this, we mixed monomeric G-actin with a kinesin-MT-PRC1 bundled network that was frozen in a particular configuration due to the lack of ATP. G-actin was polymerized over hours using a non-hydrolyzable ATP analog, AMP-PNP (29). Replacing all traces of ATP with AMP-PNP ensured the kinesin-MT network remained quiescent throughout the F-actin polymerization. Subsequently, we initiated non-equilibrium dynamics with a caged ATP precursor (30). Upon illumination with UV light, the active composite almost instantaneously transitioned from quiescence to turbulent-like flows (**SI Movie 1**). Intriguingly, the speed of the flows exhibited an initial peak, thereafter settling into a steady-state that was sustained by the ATP regeneration systems. (**Fig. 1C**). The concentration of DMNPE-caged ATP determined the speed of the kinesin motors and the overall system dynamics (11, 31). This two-step protocol was an essential prerequisite for generating reproducible dynamics. It provided ATP-independent, uniformly interdigitated F-actin/MT networks from which the non-equilibrium dynamics was initiated.

**Extensile fluid at low F-actin concentrations:** In the absence of F-actin, MTs formed a network of bundled filaments to which kinesin motor clusters bound passively. Uncaging the ATP-induced motor stepping, powering relative MT sliding. At steady state, the network rapidly reconfigured through repeating cycles of motor-driven bundle extension, buckling, fraying, and reannealing, as previously observed (11). Such network dynamics powered the large-scale flows of the aqueous solution. Motor-driven MT networks were previously studied either within viscous Newtonian fluids or passive liquid crystals (11, 23, 32). Here, we examined how a passive viscoelastic component influences the MT network dynamics.

At low F-actin concentrations, the kinesin-driven MT bundles fluidized the composite, generating an extensile fluid whose properties were tuned by the F-actin concentration. Upon initiation of activity, both the MT and F-actin networks yielded, exhibiting turbulent-like flows (**SI Movie 2, Fig. 2A**). With increasing F-actin concentrations, the flow speed decreased (**Fig. 2B**). The spatial structure of the autonomous flow fields was characterized by an exponentially decaying spatial velocity-velocity correlation. Increasing the F-actin concentration increased viscoelasticity, but had little effect on the spatial structure of the flows (**Fig. 2C**). The active flows could fluidize the F-actin network structure such that it contributed little to the dynamical length scale of the material.

The two interpenetrating networks can couple directly through steric interactions and indirectly through the flows, which are generated by the active MT network but influence both components. To investigate such coupling, we labeled both filaments. In the presence of activity, regions with high MT concentration also had higher F-actin concentrations (**SI Movie 3**). Correlating the two networks showed activity-dependent spatial colocalization (**Fig. 2D**). When active dynamics ceased due to ATP depletion, the colocalization of the two networks declined. Eventually, they became anticorrelated as the F-actin relaxed away from the static MT bundles, filling the previously empty voids (**Fig. 2D, 2E**).

**Aster formation and dynamical coexistence at intermediate F-actin concentrations:** With increasing network viscoelasticity, the flow speed decreased and approached zero for ~6 μM F-actin. Beyond this point, different dynamics emerged. In particular, the composite network separated into a high-density contracting phase which coexisted with a low-density extensile fluid (**Fig. 3A, SI Movie 4**). Initially, we observed a dense irregular system-spanning network of nodes that were connected by thin threads (**SI Movie 5**). Over time some of the threads thinned and ruptured, but asters also moved along the connecting threads to merge. Because the condensed filament-rich droplets had an outer layer enriched in MTs we refer to them as "layered asters." The droplet-like structures continued to increase in size through coalescence events, which are more indicative of liquid droplets than solid-like structures.

To quantify the emergence of structure within the active composite, we Fourier transformed the images to calculate the power spectral density (PSD) of the MT fluorescence micrographs over time. The presence of a distinct peak in the $PSD(q,t)$ is a result of a strong spatial correlation in the MT concentration, characteristic of structures with length scale, $q_m^{-1}$ (**Fig. 3B**). The position of this peak, $q_m(t)$, and its height, $PSD(q_m,t)$, changed with time. One can identify three stages of separation, delineated by the rate of change in the structural length scale. In the first stage (I), the density contrast developed between the filament-rich and filament-poor regions, leading to an increase in the magnitude of $PSD(q_m)$ but little change in the length scale, $q_m^{-1}$ (**Fig. 3C, D**). In the second stage (II), $q_m^{-1}(t)$ increased concurrently with the peak's magnitude, $PSD(q_m)$, indicating coarsening of the separated domains (**Fig. 3C, D**). The development of a peak in the *PSD* and the subsequent decrease in $q_m$ are markers of classical spinodal decomposition (33). However, in active composites $q_m$ scaled as $\sim t^{-1}$, which is much faster than the equilibrium phase separation (34-36). At the third stage (III) of separation, the rapid increase in both $PSD(q_m)$ and $q_m^{-1}$ halted and non-monotonic changes ensued, in contrast to the phase separation of passive components.

We used image segmentation to quantify the formation of filament-rich and filament-poor domains (**Fig. 3 E, F**). At the shortest times, the initially uniform composite structure became mottled with small, filament-poor voids which grew in size and number. $N_V$ is the number of voids per image frame (**Fig. 3A** at 22 min, **Fig. 3E**, right axis). Growth in $N_V$ continued in regime II, where it reached a maximum, thereafter dropping rapidly towards unity. During this transition, the many separate voids joined as the filament-rich threads ruptured. The collapse in $N_V$ captured the structural transition of the filament-rich domains from being a continuous phase to being a minority phase. There was a concomitant decrease in the area fraction of the contracted phase, $\varphi_C$, in regimes I and II (**Fig. 3E**, left axis). In stage III, $\varphi_C$ reached a minimum and began to increase, a non-monotonic behavior suggesting the emergence of new dynamics.

We examined how the material partitioned between the domain types by estimating the total F-actin and tubulin concentration from the fluorescence intensities (**Fig. 3F**). During regimes I and II, the concentration of the condensed phase, $C_C$, increased, while the concentration of the background extensile fluid, $C_F$, phase decreased. Subsequently, during stage III, $C_C$ and $C_F$, respectively, reached a peak and nadir before slowly returning to their initial values.

**Layered asters have a programmable lifetime.** We observed well-defined asters by 50 minutes (**Fig. 4A-C**). Beyond this time point, the asters began disassembling from their maximally compacted state. At the lowest F-actin concentration (8 µM), asters completely disassembled within 100 minutes (**Fig. 4a, SI Movie 6**). Above 10 µM, aster dissolution occurred over in several stages (**Fig. 4B**). First, the aster diameter slowly increased through expansion, while its filament density decreased. As asters swelled, the captured material dissociated from the asters, joining the background fluid, demonstrating that the aster's constituents were biochemically intact, but the aster stability was no longer viable (**SI Movie 6**). F-actin concentration above 30 µM stabilized the contracted asters, such that even after 5 hours little structural change occurred (**Fig. 4C, SI Movie 6**).

We quantified the aster disassembly dynamics by plotting the time evolution of the average MT concentrations within the layered asters, $I_{MT}(t)$. The MT concentration initially increased, consistent with local contractions (**Fig. 4D-F**, left axis). The concentration reached a peak and then began to decay as the asters disassembled. The decay timescale from the peak intensity varied with the F-actin concentration. For samples containing 8 µM F-actin, at 180 minutes, $I_{MT}(180)/I_{MT}(0) = 1$, indicating complete disassembly. At higher F-actin concentrations asters never fully disassembled. For 10 µM F-actin $I_{MT}(300)/I_{MT}(0)$ reached 1.2, while for 30 µM F-actin, it was 1.9. Thus, F-actin concentration governs the disassembly time scale.

The aster's effective size provided additional insight into the disassembly dynamics (**Fig. 4D-F**, right axis). At early times, the aster's average effective radius, $r_{eff}$, grew along with the intensity. In this regime, coalescence primarily drove the increase in size, while contraction drove the increase in concentration. By about 10-20 minutes, $r_{eff}$ reached a maximum and then began to decrease. $I_{MT}$ continued to increase, suggesting the first peak is driven by contraction-dominated dynamics. Subsequently, the radius began to increase with time. Thereafter, the expansion occurred as the aster intensity started dropping, a sign of the incipient disassembly. The final decay in $r_{eff}$ was not observed at 30 µM F-actin. Interestingly, the separation of these timescales was small for 8 µM F-actin, and the two peaks in $r_{eff}$ were poorly defined.

**Self-regulating structure of the layered asters**. Confocal microscopy revealed that the aster's structure consists of three concentric layers, which varied both in the composition and the local symmetry (**Fig. 5a, SI Movie 7**). The outer cortex, enriched in MTs, consisted of a monolayer of radially aligned bundles, largely excluding F-actin (**Fig. 5b**). The cortex enclosed an isotropic network of MTs and F-actin. The aster core was enriched in MTs. With increasing radius, the concentration of F-actin increased while that of MTs decreased, generating an intermediate layer enriched in F-actin.

To understand the development of the self-organized asters, we quantified the spatiotemporal organization in terms of the F-actin, MTs, and kinesin clusters within (**Fig. 6a**). We measured the average intensity, $<I_C(t)>$, for each component over time, within 10 µm of the center of each aster. The normalized F-actin concentration remained nearly constant, whereas the concentrations of the kinesin motor clusters and MTs increased (**Fig. 6b**). Notably, the increase in the motor concentration was greater than the increase in the MT concentration. The accumulation of motors over the entire aster structure was less striking (**Fig. 6c**). This highlights a preference for the motor clusters to move inwards, accumulating at the core. Such inward motion is common in aster-forming systems. It suggests that the cortex is polarity sorted with MT plus ends pointing inward (43).

Large (~100 µm) aster assemblages were propelled by the surrounding extensile fluid flow (**SI Movie 4**). When two asters encountered each other, they coalesced in a process that is reminiscent of liquid droplets, yet one that robustly preserved the aster's intricate layered structure (**Fig. 7, SI Movie 8**). The coalescence initiated as the two cortex layers overlapped. At this point, the portions of the outer cortices proximal to each other disappeared. Presumably, upon the overlap, the inactive polarity-sorted MT cortex monolayers annealed as their opposite polarity enabled efficient inter-filament sliding. At this point, the actin-rich cores of the two asters moved toward each other and joined, forming a spherical droplet-like aster. Finally, the partially disrupted outer cortex reassembled and rounded itself in a process that suggests the presence of effective surface tension.

**Bulk contraction at high concentrations of F-actin:** At intermediate concentrations of actin, the viscoelastic F-actin network quickly yielded, limiting the large-scale propagation of contractile active stresses and generating the local contraction of structured asters. At still higher F-actin concentrations, the increased network integrity permitted active stresses to percolate across the entire sample, causing the composite collapse (**Fig. 8. SI Movie 9**). After uncaging the ATP, the bulk contracting MT-actin network pulled inward, separating from the chamber margins, while maintaining the internal connectivity (**Fig. 8A**). The boundaries of the contracting structure remained uniform, while its geometry was defined by the chamber shape.

We measured the fractional width of the contracting composite over time (**Fig. 8D**). Immediately after ATP uncaging, the network exhibited little change. After a lag time of several minutes, rapid contraction ensued. The width of the contracting network was described by two exponential decays and associated timescales:

$$W(t) = a_1 e^{\frac{(t-b)}{c1}} + a_2 e^{\frac{(t-b)}{c2}} + d$$

The initial lag-time and exponential decay matched previous observations from single component filament-motor systems (15, 37). However, the double exponential form suggests a more complicated response.

We quantified the changes in the fluorescence profiles of F-actin and MTs across the sample chamber (**Fig. 8B**). Like the coexistence of structured asters and an extensile fluid, a filament-poor extensile phase coexisted with the contracting network. The MTs strongly partitioned into the contracting phase while F-actin separated more evenly. The contracting structure became heterogeneous over time. MT-poor voids developed, which yielded late-stage large variations in the MT density (**Fig. 8B**). Averaging over the local fluctuations, the density of MTs was higher at the edge of a contracting network, as seen previously (15). Interestingly, F-actin is more evenly distributed throughout the sample. At the edges of the bulk-contracted network, MT bundles extended perpendicular to the contraction direction, locally excluding F-actin (**Fig. 8C**). This outer layer is reminiscent of the MT-rich cortex of layered asters (**Fig. 5A**).

We studied how F-actin concentration affects bulk contractions. The MTs partitioned almost entirely into the contracting network while the F-actin partitioned less strongly (**Fig. 9A**). Internal fracturing played a more important role at both the lower and higher F-actin concentrations, while intermediate concentrations yielded more uniform contraction (**Fig. 9B**). MT concentration decreased as the distance from the sample edge increased. The degree to which the F-actin followed this behavior varied, suggesting that the mechanical coupling between the two networks depends on the composite's composition. For 15 µM F-actin, the density profiles of both filaments were comparable (**Fig. 9B**). At 30 µM F-actin, the profiles between F-actin and MTs were uncorrelated, and the F-actin profile did not increase at the edges. At 50 µM F-actin, there was an anti-correlation in the F-actin and MT profiles.

The final width of the contracted network exhibited a minimum for 30 µM F-actin (**Fig. 9C**). At both the low and high end of the F-actin, the contracted network started to lose its connectivity. To quantify this internal self-tearing, we used image segmentation to estimate the area fraction of MT-poor voids within the contracted domain (**Fig. 9D**). Maximal contractility yielded the smallest void fraction, while F-actin concentrations above and below that value resulted in a larger separation of F-actin and MTs. Maintaining network connectivity is important for generating maximal contractility. Another notable feature was the partitioning of the filaments between the contracting composite network and the enveloping fluid phase. Increasing F-actin concentration monotonically increased the fraction of filaments within the contracted network (**Fig. 9E**). However, the ratio of F-actin inside and outside the contracted network decreased as a function of F-actin concentration and approached unity (inset **Fig. 9E**). These observations suggest that the composite network changes from a low-density limit where both filaments co-contract, to a high-density limit where the MT-network contracts within a viscoelastic suspension of entangled F-actin.

**State diagram of the actin-MT composites**: Our work builds on the previous studies of active composites (38, 39). We have described how increasing F-actin concentrations induce transitions from the extensile fluid to layered asters, to global contractions. To map out the entire phase diagram we varied both the F-actin and the ATP concentration. Because the dynamics were time-dependent, the diagram summarizes structures observed after one hour (**Fig. 10**). Low ATP concentrations drove contractions, even in the absence of F-actin. This echoes previous results which established extensile MT networks as yield stress solids, where active stresses fluidized the network (23, 40). While F-actin concentration determines the composite's viscoelasticity, the ATP concentration plays a dual role: it controls both the active stresses and the sample viscoelasticity. Kinesin-generated active stress is determined by its ATP-dependent stepping rate (41). Between the sequential force-generating steps that kinesin motors take, they remain attached to the MT, acting as passive crosslinkers. Lower ATP concentrations increase the fraction of kinesin clusters that passively cross-link MTs. Thus, decreasing ATP effectively increases the viscoelasticity of the MT network alone (23). Consequently, it is not surprising that ATP concentration, like F-actin, also plays a role in determining the composite's behavior at non-zero F-actin concentrations. Similar to F-actin, increasing ATP concentration induced transitions from the bulk contraction to structured asters and extensile fluids.

**Extensile bundles drive bulk contractions:** The microscopic details that underlie the emergence of the above-described dynamics remain unclear. To gain insight, we quantified the behavior of individual MT bundles within the contracting network. Isolated fluorescent MT bundles were doped into a network of unlabeled filaments. Upon initiation of activity, these bundles extended to several times their initial contour length before buckling (**Fig. 11**). Therefore, these results demonstrate that, despite the long-term contraction, the initial kinesin-driven activity generates MT-bundle extension. Such observations provide a glimpse into the microscopic mechanisms by which active networks produce two seemingly separate processes of extensile fluid flows and solid-like contractions.

**Discussion:** The dynamics of active matter is determined by the balance between active stresses, generated by the motile constituents, and the viscous or elastic reaction stresses that arise due to the deformation of the material. In single-component formulations of cytoskeletal-based active matter, independent control of active and passive stresses can be challenging. For example, in conventional MT-based active fluids, kinesin motors act both as passive crosslinkers and generators of active forces (23, 42). Consequently, decreasing the ATP concentration decreases the magnitude of the active stresses while increasing the elastic modulus of the material. This dual role limits the full exploration of the non-equilibrium phase space, while also impeding comparison to theory. The MT-actin active composites overcome these obstacles, allowing one to independently tune both the active stresses that are exclusively generated by the MT component and the passive stresses that are largely determined by the viscoelastic F-actin network. Separating these two control parameters yielded several unexpected findings. First, increasing the passive viscoelasticity changed the symmetry of the self-organized active stresses, inducing a transition from extensile to contractile dynamics. Second, the transition from extensile to contractile stresses was marked by the coexistence of two dynamically organized structures: filament-rich layered asters and filament-poor extensile fluids. Third, the layered asters exhibited self-regulating spatial organization. Finally, they were also transients whose lifetimes were encoded in the sample viscoelasticity. We discuss each of these observations.

*Passive viscoelasticity controls the symmetry of active stresses:* With notable exceptions from 2D motility assay geometries (43, 44), cytoskeletal active matter typically exhibits either uniform contractile or extensile dynamics, which implies the existence of a single type of self-organized active stress (13, 16, 39, 45). It remains a challenge to predict the nature of the mesoscopic active stresses given known microscopic dynamics. Several mechanisms have been proposed including: (1) the spatial arrangement of motors along polar filaments and their enhanced propensity for tip binding, (2) the spatially dependent dynamics of motor-stepping along the filaments, and (3) the asymmetry introduced by mechanical rigidity of the constituent filaments (17, 46, 47). Disentangling these mechanisms in a single component network is challenging as filament spatial organization and mechanics, and the dynamics of motors on those filaments, are interrelated (48). We found that a viscoelastic F-actin network can induce a transition from extensile flow to bulk contractions. We also observed the coexistence of extensile to a contractile state. These dual observations challenge existing theoretical models. They are also in contrast to previous work with myosin-driven MT-actin composites, which showed that the viscoelastic response could be tuned by the addition of a second material component, but the symmetry of the active stress did not change (39). Even in contractile conditions, the constituent MT bundles initially exhibited extensile dynamics (**Fig. 11**). Unfortunately, we were unable to probe their microscopic dynamics at longer times. Notably, the rectification of extensile into contractile forces was demonstrated for non-linear elastic networks (49). The applicability of these descriptions to the viscoelastic active composites studied here requires further investigation.

We speculate about a possible mechanism by which the viscoelasticity controls the symmetries of the active stresses. Steady-state flows of extensile fluids are driven by a cyclical process of motor-driven MT bundle extension, buckling, fraying, and reannealing. The extension requires bundles with mixed polarity, in which the MT plus ends are uniformly distributed and are equally likely to point in both directions. Molecular motors quickly polarity sort an isolated bundle, yielding static spatial domains, within which all MT plus ends point in the same direction. It is assumed that the network reconfiguration dynamics continuously mix the polar domains with the MT bundles, which in turn sustains the steady-state generation of the extensile stresses and the associated non-equilibrium turbulent dynamics. It is possible that the viscoelastic actin network sterically suppresses the buckling and the reannealing of MT bundles. The suppression of such dynamical modes could quickly generate fragmented and polarity sorted MTs domains. Such architectures can no longer sustain extensile stresses. However, a microtubule network could exhibit contractile dynamics through end-binding or some other mechanisms.

*Active contractile dynamics generate layered aster composites:* We also showed that the extensile to contractile transition is associated with the formation of layered asters. During the initial stages of the separation, we observed thin system-spanning threads which yield and rearrange to form spherical filament-rich asters. Similar morphological evolution is observed in the passive phase separating viscoelastic polymer mixtures, but in this case, each phase is enriched in one polymer type (33, 50). In the active system, both filaments separated into one phase whose composition was not governed by thermodynamics. On longer time scales, we observed the slow differential motion of the various components driving the onion-like layering. (**Fig. 6**). In contrast, previous experiments with myosin-driven MT-actin composites exhibited purely contractile behavior in which the actin and MT movement were indistinguishable (39).

Simple asters have been studied previously (51-55). Their formation and stability are explained by invoking the preference of molecular motors to reside at a filament tip (56). Aster radial size is comparable to the filament length and the radial organization extends from the motor-rich core to their periphery. In contrast, layered asters have an outer cortex that is many times the length of the constituent MTs. Furthermore, the cortex envelops a sizeable region of isotropically organized filaments. The mechanisms that determine the size of the cortex and the asters' actin-rich core remain unknown.

Structured asters and bulk contracted composites share certain structural features. Notably, bulk contractions yield a monolayer of dense MTs which emerge at the very edge of the contracting structure. This layer is reminiscent of the MT-rich cortex observed in layered asters. Both layers exclude F-actin while the MT bundles point along the surface normal. Similar edge-bound structures have also been observed in other contractile systems (57, 58). Thus, they might be a ubiquitous, yet still poorly understood, a feature of bulk contractions. Understanding the molecular mechanisms that lead to their formation, might provide insight into the formation of layered asters.

*Structured asters have a finite lifetime:* Another poorly understood feature of layered asters is their finite and programmable lifetime. The intrinsic length and time scales of kinesin-1 motors moving on MTs are microns and seconds (41, 59). By comparison, the asters remain stable for tens of minutes to hours. Imbuing synthetic self-assembly processes with a finite lifetime remains a challenge, especially when compared to cells that rapidly assemble large-scale structures, such as a mitotic spindle or a cytokinetic ring, at a predetermined location for a predetermined time, before equally rapidly disassembling them. Understanding the mechanisms that govern the asters' finite lifetimes might reveal strategies with broader applications. One can only speculate about a possible mechanism that endows structured asters with a finite lifetime. The MT-cortices likely have polar order. Thus, they can act as a highway for motor clusters to move inward, accumulating in the aster core. The buildup of motor proteins beyond a critical concentration could destabilize the aster. Furthermore, by linking multiple MTs, the inward moving motor clusters could contribute to the stability of the outer monolayer cortex and the entire structured aster. Exploring the role of PRC1 crosslinkers, which specifically bundles antiparallel MTs, in the stability of the asters and the outer MT cortex, remains a topic for future exploration.

More broadly, layered asters, which minimize their surface area and coalesce with each other, are reminiscent of liquid droplets. Liquid-liquid phase separation (LLPS) has emerged as a key organizing principle in cell biology (60, 61). Thus, our observations raise intriguing questions at the nexus of the emerging areas of active matter, phase separation in cell biology, and biophysics of cytoskeleton. From the perspective of cell biology, our findings should promote studies if and how a combination of cytoskeletal fibers and non-equilibrium fluctuations can promote LLPS(62). From the active matter perspective, some observed phenomenology is reminiscent of viscoelastic phase separation observed in the passive binary polymer mixtures. Further experiments are necessary to determine if there are any fundamental similarities between our observations and the classical phase separation.

Our work revealed the formation of intricate spatiotemporal patterns through mechanical interactions between actin filaments and MTs alone. Understanding the mechanism governing our simplified systems might shed insight into several phenomena in cell biology, which involve interplay between MTs and the actin cytoskeleton. Amongst others, these include MT penetration into filopodia, lamellipodia, and growth cones, MT 'tracking' along the actin stress fibers, meiotic MT spindles that are surrounded by and embedded in the actin network, MTs growing out of centrosomes, which are also actin-organizing centers, and the intricate interplay of actin and MT cytoskeleton that controls the onset of cytoplasmic streaming in Drosophila oocyte (63-67). In the last example, the onset of cytoplasmic streaming is determined by depolymerization of the viscoelastic actin network. Intriguingly, this is a ubiquitous feature of reconstituted active/MT composites studied here.

In summary, we showed that MT-actin composites exhibit distinct organized states that evolve. The coexisting dynamical structures are reminiscent of the multiple physiological states observed in cell biology. Elucidating the governing equation of layered asters and the extensile fluids represents a challenge for existing theoretical formalisms of active matter. Answering this challenge might enable real-time control of coexisting out-of-equilibrium states, a hallmark of dynamical and adaptable biological cells.

**Materials and Methods**

**MTs:** Short, fluorescently labeled GMPCPP stabilized MTs were generated by established methods (51). Briefly, recycled Alexa-647 labeled and unlabeled tubulin monomers were polymerized in excess GMPCPP to yield short, fluorescent MTs with a labeling efficiency of about 1%. Small aliquots were snap-frozen in 80 mM PIPES, 2 mM magnesium chloride, pH 7.4 (M2B) at 8 mg/mL, and used throughout the experiment.

**Kinesin Motor Clusters:** Short biotinylated kinesin constructs were prepared via established protocols (11). Simply, K-401-BCCP was expressed in E. coli cells, harvested, and purified via HPLC. These biotinylated motors were snap-frozen in M2B and used throughout the experiment. To generate motor clusters, the biotinylated motors were incubated in a 2:1 ratio with streptavidin tetramers. In the case of fluorescently labeled motor clusters, Dylight 488 neutravidin was used in place of streptavidin.

**PRC1-NSΔC:** PRC1-NSΔC, a truncated version of the MT crosslinker, PRC1, was expressed in and purified from E. Coli. The non-structured region at the C-terminus of the PRC1 protein is deleted which reduces the interactions between adjacent crosslinkers. These proteins were snap-frozen in 50 mM sodium phosphate buffer (pH 7.0), 400 mM imidazole, 150 mM sodium chloride, 5 mM TCEP, and 2 mM EDTA as described previously (27).

**Actin:** Rabbit muscle actin was purified as previously described (68). The nucleotides in the G-actin solution were then replaced by AMP-PNP using previous protocols (69). Briefly, G-actin solution was passed through ion exchange resin to bind free nucleotides directly into buffers prepared with AMP-PNP, a non-hydrolyzable ATP analog. G-actin solutions (~150 µM) were snap-frozen in 5 mM Tris-HCl (pH 8.0), 0.2 mM $CaCl_2$, 1 mM AMPPNP, and 0.5 mM DTT.

**Chambers:** Flow cells in which the experiment took place were created by sandwiching Parafilm between acrylamide brush coated glass slide and coverslips. The Parafilm is then briefly raised above the melting temperature, creating a sealed chamber with inlets, outlets, and a size of about 10 mm x 30 mm x 0.1 mm. After loading the experimental solutions, the inlet and outlet are sealed with silicone vacuum grease.

**Experiment:** A bundled MT network was prepared in solution with actin monomers, loaded into chambers, sealed, and polymerized over 2 hours in darkness. Then, UV light was shown on the sample, releasing caged ATP and allowing kinesin activity to commence. The bundled MT network was formed by combining formed GMPCPP MTs, PRC1-NSΔC, kinesin motor clusters, and caged ATP under red LED light. This solution also contains an ATP regeneration system (21 units/mL pyruvate kinase and 40 mM PEP) and anti-oxidants (18.7 mM glucose, 1.4 µM glucose oxidase, and 0.17 µM catalase). To this, the appropriate amount of actin is added, mixed, and loaded into the experimental chamber, which is promptly sealed. After two hours of actin polymerization, activity is initiated by exposing the sample to UV light (365 nm) for 30 seconds. The experimental slides can hold up to 12 parallel chambers, allowing for experiments to probe the effects of the varied parameter (e.g. actin concentration) on the same experimental preparation.

**Microscopes:** The samples are observed on a Nikon Ti microscope with 4x, 10x, and 20x objectives, Andor Neo or FLIR Blackfly cameras controlled by Micromanager. Confocal microscopy was performed using a Leica SP8 confocal microscope.

**Acknowledgments:** We thank Bulbul Chakraborty, Mike Norton Moumita Das, Margaret Gardel, Chase Broedersz, Andreas Bausch, Gijsje Koenderink, Aparna Baskaran, and Michael Norton for illuminating discussions. We thank Radhika Subramanian for the PRC1-NSΔC construct. This work was supported by the Brandeis NSF-MRSEC, DMR-2011486. We also acknowledge the use of MRSEC Optical Microscopy and Biological Synthesis facilities supported by the grant DMR-2011486.

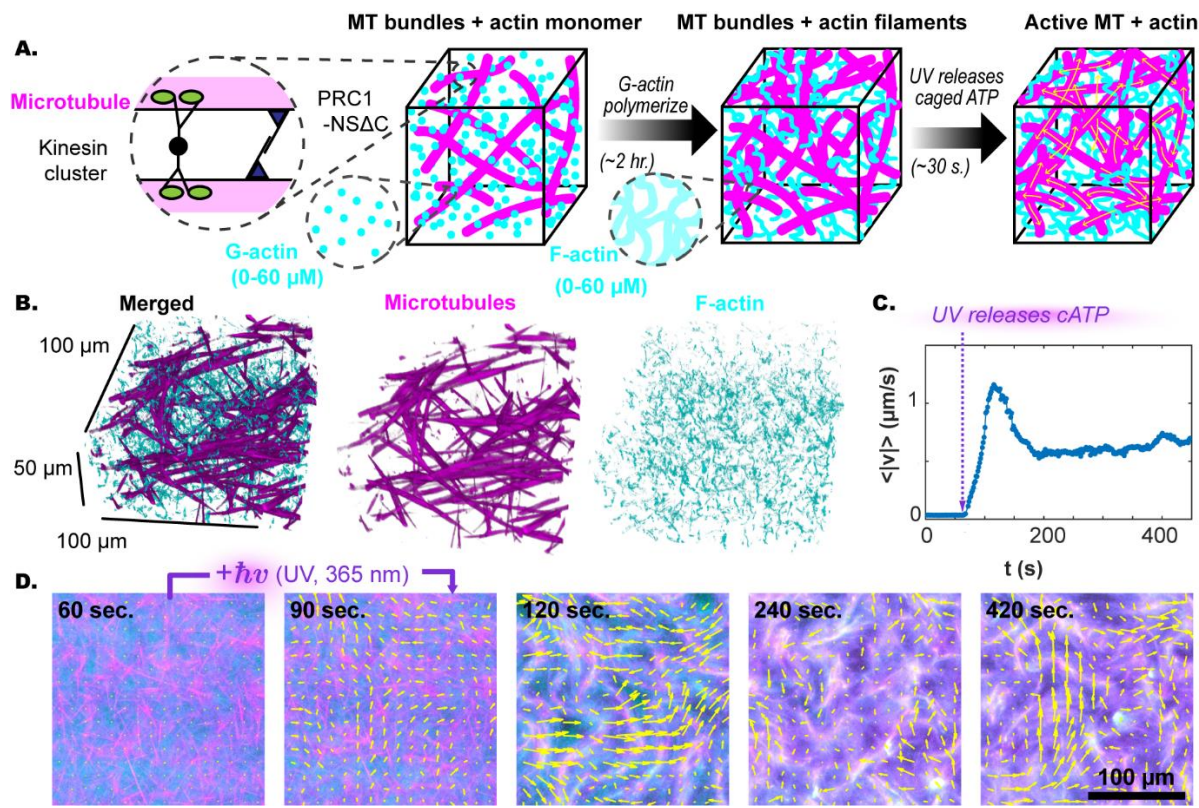

**Figure 1. Light-activated F-actin/MTs composites**. **(A)** GMPCPP-stabilized MTs are bundled by a specific MT crosslinker, PRC1-NSΔC, and driven by streptavidin-based kinesin motor clusters. AMP-PNP-bound G-actin is added to the MT bundle network and polymerizes in the high-salt buffer. After two hours,

filamentous actin formed a uniform interpenetrating actin network. At that point, DMNPE-caged ATP was cleaved with UV light, and the kinesin motors generated large-scale reorganization. **(B)** Confocal microscopy shows the isotropic uniform interpenetrating network of actin and MTs. **(C)** UV illumination releases caged-ATP, generating flows measured by PIV. **(D)** Kinesin-driven flow drives the re-organization of the actin and MT networks. Yellow arrows indicate the velocity field.

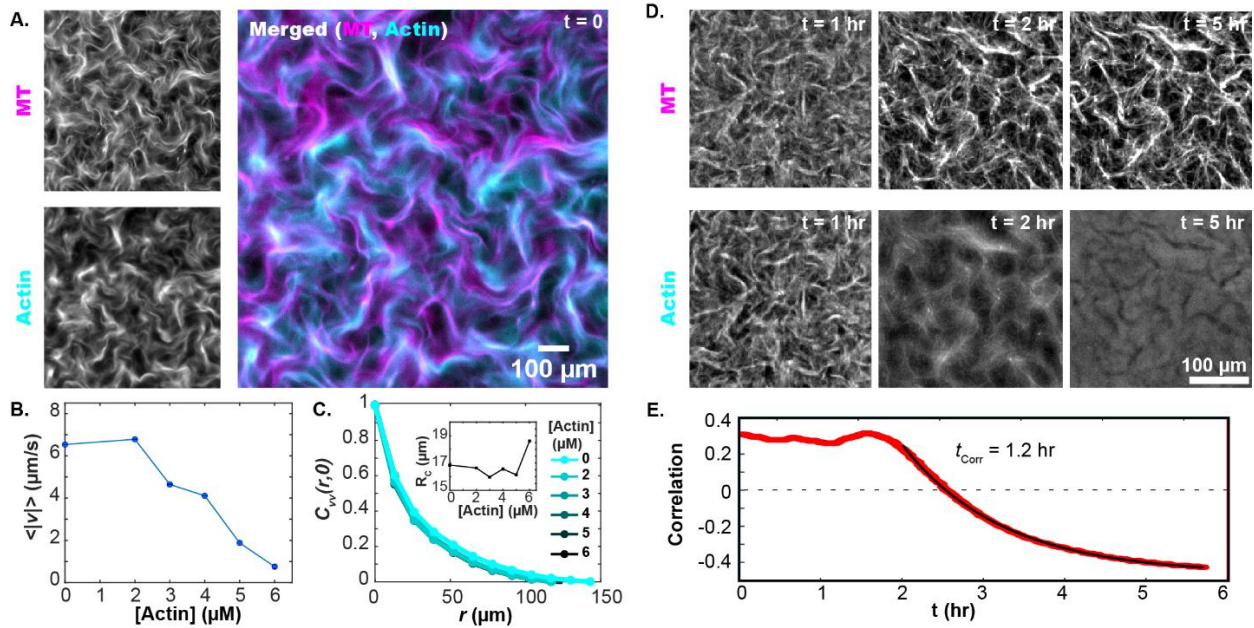

**Figure 2. F-actin controls the dynamics of the extensile MT network.** **(A)** Kinesin motors drive the reorganization of both F-actin and MT networks. **(B)** The mean speed of the autonomous flows decreases with increasing F-actin concentration. **(C)** The spatial velocity-velocity correlation function of the autonomous flows is independent of the F-actin concentration. **(D)** Upon cessation of motor-driven flow, F-actin relaxes away from its out-of-equilibrium configuration (red curve). The time evolution of this viscoelastic relaxation is distinct from the timescale of the active network halting (blue curve) **(E)** Spatial correlation between MT and F-actin concentration field as a function of time.

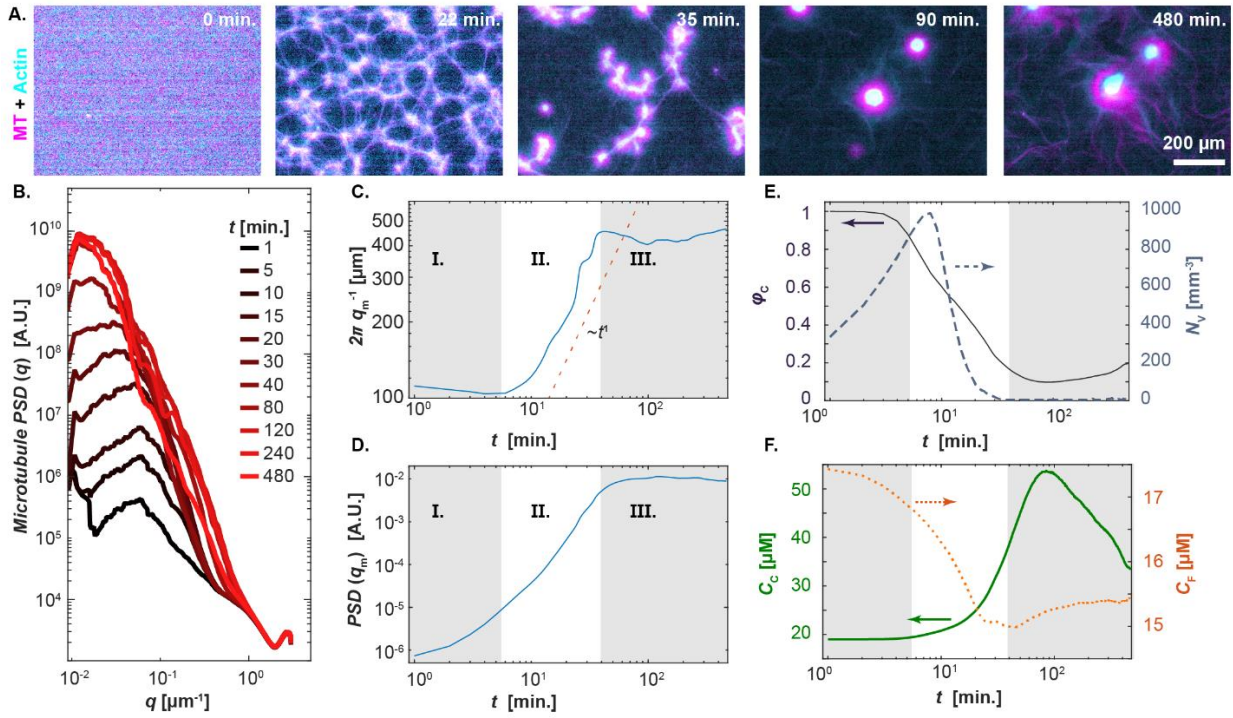

**Figure 3. Structure of self-organized layered asters**. **(A)** Time series showing a uniform composite phase separating into filament-rich aster and filament-poor extensile fluids. **(B)** The power spectral density, PSD($q$), of the MT fluorescence micrographs shows the emergence of a peak with increasing time. **(C)** The position of the PSD peak density, $q_m$, plotted against time exhibits three stages of phase separation. **(D)** The magnitude of the PSD peak is plotted as a function of time. **(E)** Time-dependent fractional area of the contractile domain (solid line, left axis) and the number density of the interstitial extensile fluid domains (dashed line, right axis) extracted from real-space images. **(F.)** The combined F-actin and MT concentration within the contractile aster phase (solid line, left axis) and active fluid phase (dotted line, right axis) as a function of time. Samples contained: 100 µM cATP, 1.3 mg/mL MTs, and 3 µM actin.

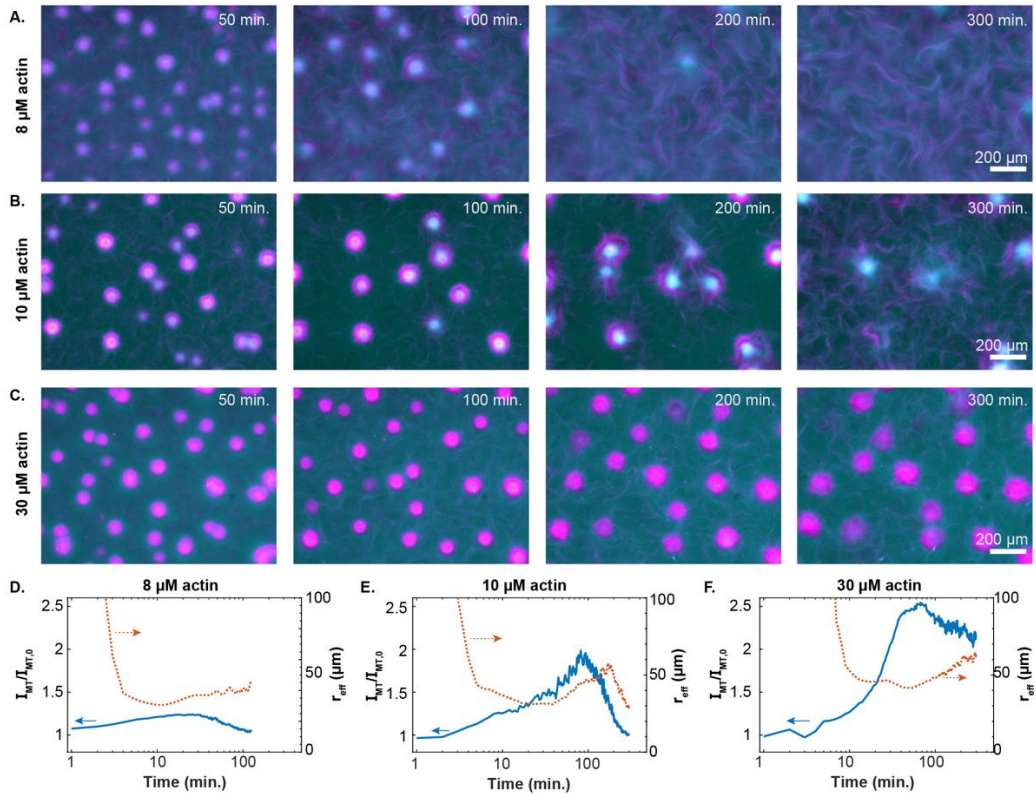

**Figure 4. Programmable finite lifetime of layers asters:** **(A-C)** Time-lapse images show the structural changes of contracted asters embedded in an active fluid at 8, 10, and 30 µM F-actin. **(D-F)** Normalized MT fluorescence intensity within the contracted aster phase (solid line, left axes) and equivalent spherical radius of aster domains (dashed line, right axes) show temporal evolution depends on actin concentration (8, 10, and 30 µM actin).

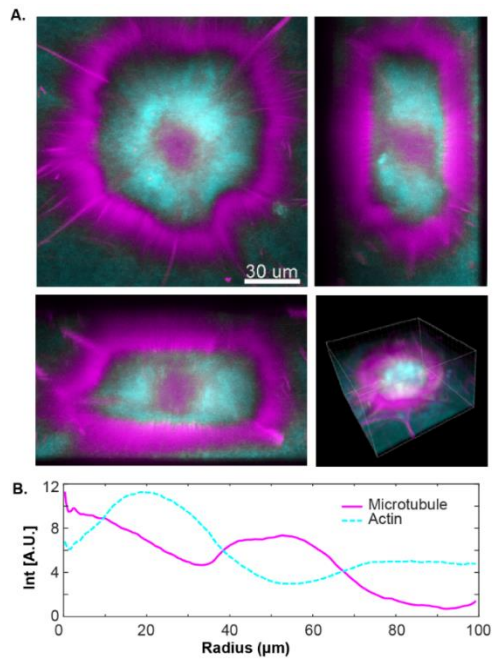

**Figure 5. Structure of layered asters**. **(A)** Two-dimensional projections of an aster imaged with confocal microscopy. **(B)** The radial profile of MTs and actin reveals a layered organization consisting of an MT-rich core, an actin-rich intermediate layer, and the outer cortex composed of radially aligned MTs.

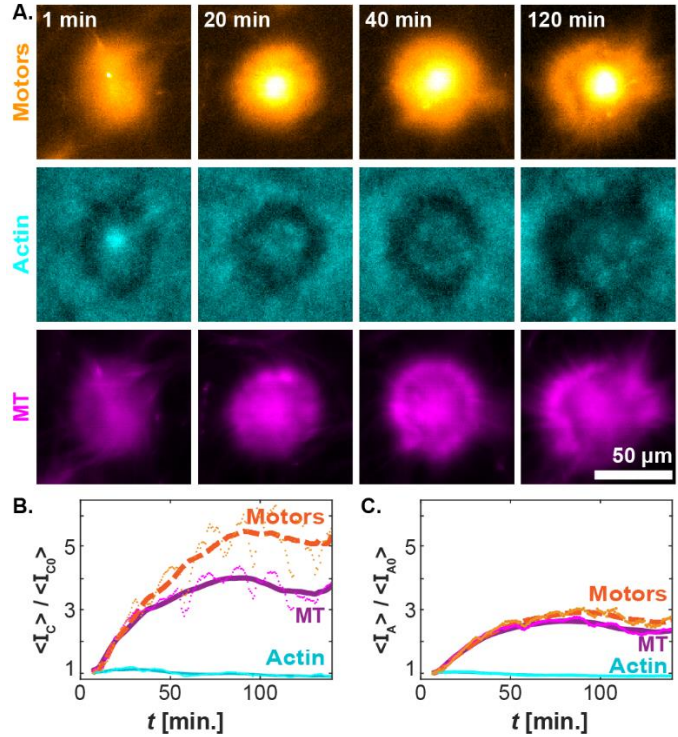

**Figure 6. Temporal evolution of the aster structures.** **(A)** Time-lapse of a single aster for the motor (top), actin (middle), and MT (bottom) channels. **(B)** Time evolution of the normalized mean intensity within 6.5 µm of the centroid. **(C)** Time evolution of normalized intensity over entire contracted domain Lines represent a moving average over 20 minutes. Data from 300 µM caged ATP, 6 µM actin, 100 µM caged ATP.

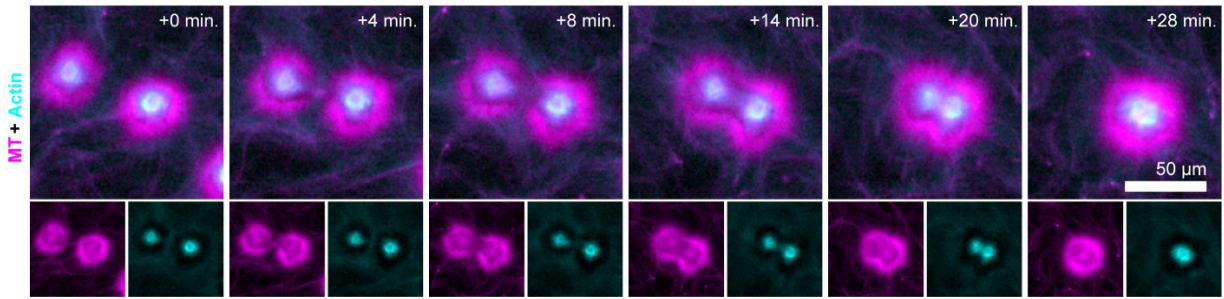

**Figure 7. Liquid-like coalescence of layered asters.** Time-lapse images showing the liquid-like coalescence of two nearby asters. Coalescence events preserve the intricate structure of the layered asters.

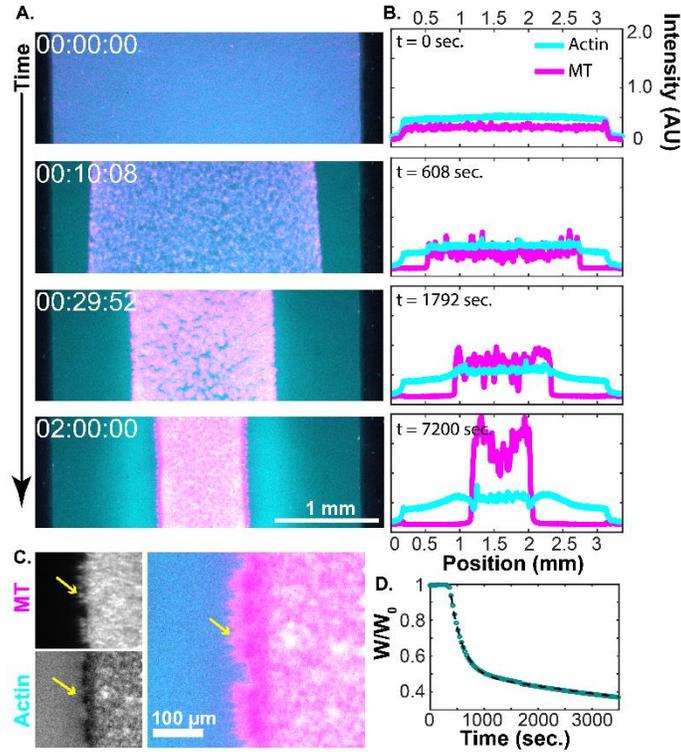

**Figure 8. Bulk contraction at high F-actin concentrations.** (**A**) Dual-color fluorescence micrographs of actin (cyan) and MTs (magenta) show the MT-rich network contract towards the chamber interior. (**B**) Intensity profiles of fluorescence micrographs show the accumulation of F-actin and MTs during bulk contraction. While nearly all of the MTs are carried in the contracting structure, a significant fraction of F-actin is left behind. (**C**) MT bundles point outward from the contracted domain. (**D**) A plot of the normalized width as a function of time. The fit to Eq. 1 yields $c_1$=203 s and $c_2$=3470 s and the lag time $b$=269 s.

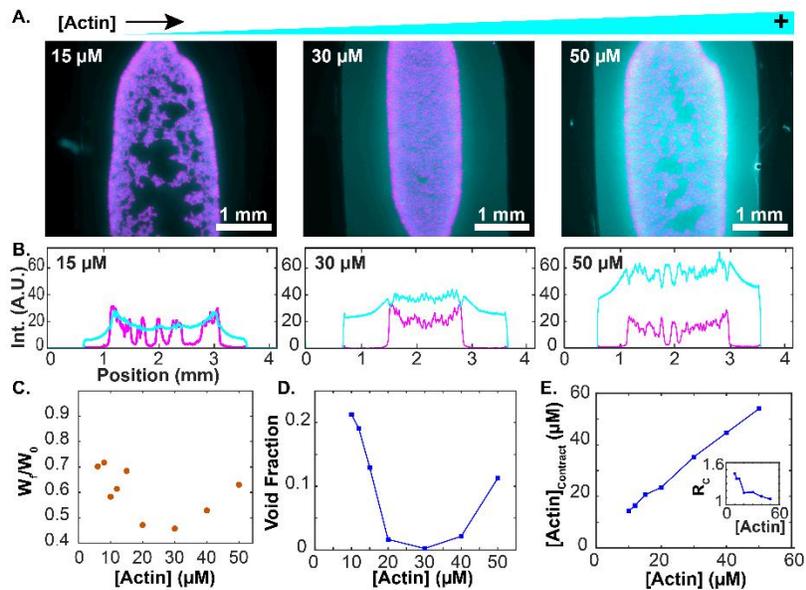

**Figure 9. Bulk contraction changes non-monotonically with F-actin concentration.** (**A**) Fluorescence micrographs of bulk-contracted domains at various F-actin concentrations. Images represent the densest

structures formed. **(B)** Profiles of the intensity of the actin and MT signals across the sample chamber. **(C)** The normalized width of the final contracted structures varies non-monotonically with the F-actin concentration. **(D)** The void fraction of MTs within the contracted structures depends on the F-actin concentration. **(E)** Increasing overall F-actin concentration increases the concentration of F-actin within the contracted domains. Inset: The ratio of F-actin concentration in and outside the contracted domains.

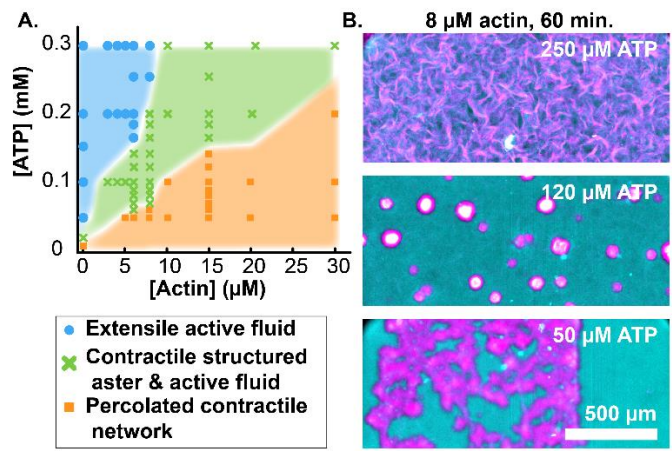

**Figure 10. Dynamics phase diagram of actin/MT active composites. (A)** State diagram representing the self-organized structure observed after one hour of activity as a function of ATP and F-actin concentration. Bulk contraction is characterized by the retraction of the sample edges before self-tearing into localized structures. **(B)** Representative merged fluorescence micrographs of the active composite demonstrating self-organized structure at fixed actin concentration while varying ATP concentration (MT, magenta; F-actin, cyan).

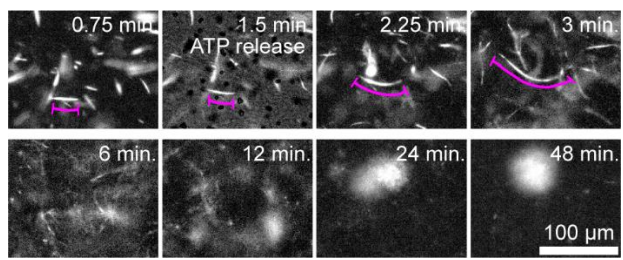

**Figure 11**. **Bulk contractions are driven by extensile MT bundles:** Time-lapse images showing initial extension of fluorescently labeled MTs in a composite network. At longer times, bulk contraction is observed. 100 µM caged ATP, 8 µM actin.